\documentclass[10pt,fleqn]{article}
\usepackage{}  
\usepackage{CJK}   
\usepackage{indentfirst}  
\usepackage{bm}    
\usepackage{graphicx}  
\usepackage{subfigure}
\usepackage{cite}
\usepackage{slashbox}

\begin{CJK*}{GBK}{song}
\begin{document}



\begin{center}
\LARGE\bf Spinning Particle in Gravitational Field of Black Hole Involving Global Monopole
\end{center}

\footnotetext{\hspace*{-.45cm}\footnotesize $^*$ Corresponding author }

\begin{center}
\rm Qihong Huang, \ \  Juhua Chen \ and  \ Yongjiu Wang$^{*}$
\end{center}

\begin{center}
\begin{footnotesize} \sl
College of Physics and Information Science, Hunan Normal University, Changsha, Hunan 410081

E-mail: wyj@hunnu.edu.cn

\end{footnotesize}
\end{center}


\begin{center}
\begin{minipage}{15.5cm}
\parindent 20pt\footnotesize
In this paper we study the dynamics of trajectory of a spinning particle in a Schwarzschild spacetime involving a global monopole. We set up the equations of motion and find three types of trajectories. We study the conditions that a spinning particle, originally moving in the innermost stable circular orbit around the black hole involving a global monopole, will escape to infinity after it is kicked by another particle or photon. Three types of trajectories of a spinning particle in a Schwarzschild spacetime involving a global monopole are simulated in detail and  the escaping energy and velocity of the spinning particle is also obtained in the present paper.
\end{minipage}
\end{center}

\begin{center}
\begin{minipage}{15.5cm}
\end{minipage}
\end{center}

\section{Introduction}
It is an interesting topic to study the motion of the spinning particle, such as Dirac fermions, in curved spacetime by using pseudo-classical mechanics model, in which the spin degrees of freedom of the particle are described with anti-commuting Grassmann variables\cite{Berezin,2,3,4,5,6,7,8,9}. The equations describing the motion of spinning particle were derived firstly by Papapetrou\cite{f1} and were reformulated by Dixon\cite{f2}. An alternative set of equations were obtained by van Holten\cite{f3}. When the four-momentum and four-velocity of the spinning particles become co-linear, the Dixon-Souriau equations\cite{f4} reduce to the van Holten equations\cite{f5}. Rietdijk and van Holten\cite{10} analyzed the motion of spinning particle in the Schwarzschild spacetime and they derived an exact precession equation of the perihelion for the planar orbit. Suzuki and Maeda\cite{g1} had studied the chaotic motion of a spinning test particle in Schwarzschild spacetime with the Poincar\'{e} map and the Lyapunov exponent. Verhaaren and Hirschmann\cite{g2} had found the chaotic orbits for spinning particle with small spin values in Schwarzschild spacetime and developed a new method by comparing the Lyapunov exponent of chaotic orbits. Han\cite{g3} had numerically studied the chaotic dynamics of spinning particle in Kerr spacetime with Papapetrou equations. Hossain\cite{g4} had studied the geodesic motion of pseudo-classical spinning particle in the Reissner-Nordstr\"{o}m-de Sitter spacetime and discussed its bound state orbits. Chen {\sl et al} had investigated the geodesic structure of test particle in several black hole spacetimes\cite{chen1,chen2}. Wang and Wu\cite{ww} had investigated the contributions of the next-order spin-orbit to chaos in spinning compact binary system by numerically individual orbit simulations. The corresponding chaos phenomena about the spinning compact binary system were discussed in Ref.\cite{wx1,wx2,zw}. The mechanism of jet formation of the black hole is one of the most intriguing and important problem in modern astrophysics. Al Zahrani, Frolov and Shoom\cite{14} had presented a jet model of charged particle which moves around a weakly magnetized Schwarzschild black hole.

Monopole, as a result of gauge-symmetry breaking, is similar to the elementary particle and has Goldstone field with the energy density decreasing with $r^2$. The large energy in the Goldstone field surrounding the global monopole suggests that they can produce a strong gravitational field. Researching on monopole is important for the topological defect in the early universe and other physical effects\cite{a1,a2,a3,a4,a5}. When the Schwarzschild black hole swallows a global monopole, it will form a black hole involving a global monopole system and possessing a solid deficit angle, which is different from the Schwarzschild black hole alone.

In this paper, we plan to study the motion of spinning particle in the gravitational field of black hole involving a global monopole. When the global monopole is absent, our result will reduce to the Schwarzschild case\cite{10}. The organization of this paper is as follows. In Section 2 we briefly review the relevant equations and constants for the motion of spinning particles in curved spacetime, and investigate the motion of spinning particles in the gravitational field of black hole involving a global monopole. In Section 3 we study the conditions that a spinning particle, originally revolving around the black hole involving global monopole in the innermost stable circular orbit, will escape to infinity after it is kicked by another particle or photon. The critical escape energy and velocity of the particle is obtained. In last section we present our brief remarks. Throughout the paper we choose units $\hbar=c=G=1$.

\section{Equations of motion for a spinning particle in gravitational field of black hole involving global monopole}
The geodesics of the spinning space is given by the action
\begin{equation}
S=m\int_1^2 {d\tau}\bigg(\frac{1}{2}{g_{\mu\nu}(x)}{\dot{x}^\mu}{\dot{x}^\nu}+\frac{i}{2}{g_{\mu\nu}(x)}{\psi^\mu}{\frac{D\psi^\nu}{D\tau}}\bigg),
\end{equation}
where $m$ is the dimension of mass and $g_{\mu\nu}$ describes the metric of the background spacetime. The covariant derivative of the Grassmann coordinates $\psi^\mu$ is defined by
\begin{equation}
\frac{D\psi^\mu}{D\tau}={\dot{\psi}^\mu}+{\dot{x}^\lambda}{{\Gamma}_{\lambda \nu}^\mu}{\psi^\nu}.
\end{equation}
The overdot denotes an ordinary derivative about proper time $\frac{d}{d\tau}$. The trajectory, which makes the action stationary under variations vanishing at the endpoint, can be cast in the follow form
\begin{eqnarray}
\frac{{D^2}x^\mu}{D{\tau}^2}=\ddot{x}^\mu+{{\Gamma}_{\lambda \nu}^\mu}{\dot{x}^\lambda}{\dot{x}^\nu}={\frac{1}{2i}}{\psi^\kappa}{\psi^\lambda}{{{R_{\kappa \lambda}}^\mu}_\nu}{\dot{x}^\nu},\qquad \frac{D\psi^\mu}{D\tau}=0.
\end{eqnarray}

The anti-symmetric tensors, which describe the relativistic spin particle and can formally be regarded as the spin-polarization tensor, are given by
\begin{equation}
S^{\mu\nu}=-i{\psi^\mu}{\psi^\nu}.
\end{equation}

The first part of Eq.(3) implies the existence of a spin-dependent gravitational force , while the second part asserts that the spin is covariant constant
\begin{equation}
\frac{DS^{\mu\nu}}{D\tau}=0.
\end{equation}

The space-like components $S^{ij}$ are proportional to the magnetic dipole moment of the particle. The time-like components $S^{i0}$ represent the electric dipole moment which vanish for free Dirac particle, such as free electron and quark, in the rest frame. In the Grassmann coordinates,  a covariant constraint is
\begin{equation}
{g_{\mu\nu}(x)}{\dot{x}^\mu}{\psi^\nu}=0.
\end{equation}
When the action Eq.(1) is invariant under the transformations $\delta{x^\mu}={R^\mu(x,\psi)}$ and $\delta{{\psi}^\mu}={S^\mu(x,\psi)}$, there exists a constant of motion
\begin{equation}
J^{(\alpha)}(x,p,\psi)=B^{(\alpha)}(x,\psi)+{p_\mu}{R^{(\alpha)\mu}(x,\psi)},\qquad p_\mu=mg_{\mu\nu}{\dot{x}^\nu},
\end{equation}
which asserts that the Killing vector $R^{(\alpha)\mu}$ gives a contribution to the orbital momentum and the contribution of spin is contained in the Killing scalars $B(x,\psi)$. Without the Killing scalars $B(x,\psi)$, the Killing vector does not give a conserved quantity of motion. The Killing scalars $B(x,\psi)$ is given by
\begin{equation}
B^{(\alpha)}_{,\mu}+\frac{\partial B^{(\alpha)}}{\partial \psi^\lambda}{\Gamma^\lambda_{\mu\kappa}}{\psi^\kappa}=\frac{i}{2}{\psi^\kappa}{\psi^\lambda}{R_{\kappa\lambda\nu\mu}}{R^{(\alpha)\nu}},
\end{equation}
which shows that each Killing vector $R^{(\alpha)\nu}$ is associated with a Killing scalar $B^{(\alpha)}$. The quantity $R_{\kappa\lambda\nu\mu}$ denotes the Riemann curvature. The Killing vectors $R^{(\alpha)\nu}$ satisfy the equations
\begin{equation}
R^{(\alpha)}_{\mu;\nu}+R^{(\alpha)}_{\nu;\mu}+\frac{\partial {R^{(\alpha)}_\mu}}{\partial \psi^\lambda}{\Gamma^\lambda_{\nu\kappa}}{\psi^\kappa}+\frac{\partial {R^{(\alpha)}_\nu}}{\partial \psi^\lambda}{\Gamma^\lambda_{\mu\kappa}}{\psi^\kappa}=0.
\end{equation}

The symmetry of the spinning particle model can be divided into two classes. First, there are constants of motion that exist in any theory and these are called "generic". The second kind of constants of motion depend on the specific form of the metric $g_{\mu\nu}(x)$ and they are called "nongeneric". For the spinning particle model defined by the action (1), there are four generic constants of motion\cite{11}.

The two most important and obvious ones are the world-line Hamiltonian
\begin{equation}
H=\frac{1}{2m}g^{\mu\nu}(x)p_\mu p_\nu
\end{equation}
and the supercharge
\begin{equation}
Q=p_\mu\psi^\mu,
\end{equation}
the other two generic constants of motion are the dual supercharge
\begin{equation}
Q^*=\frac{1}{3!}\sqrt{-g}\varepsilon_{\mu\nu\kappa\lambda}p^\mu\psi^\nu\psi^\kappa\psi^\lambda
\end{equation}
and the chiral charge
\begin{equation}
\Gamma_*=\frac{1}{4!}\sqrt{-g}\varepsilon_{\mu\nu\kappa\lambda}\psi^\mu\psi^\nu\psi^\kappa\psi^\lambda.
\end{equation}
Condition (6) implies that
\begin{equation}
Q=0.
\end{equation}

The gravitational field of black hole with a global monopole is described by the line element\cite{12}
\begin{equation}
ds^2=-\bigg(1-8\pi\eta^2-\frac{\alpha}{r}\bigg)dt^2+\bigg(1-8\pi\eta^2-\frac{\alpha}{r}\bigg)^{-1}dr^2+r^2d\theta^2+r^2\sin^2\theta d\varphi^2,
\end{equation}
where $\alpha=2M$, $M$ the total mass of black hole and $\eta$ is the scale of the symmetry breaking. For a typical grand unification scale $\eta\sim10^16GeV$, we have $8\pi\eta^2\ll1$. The corresponding Killing vector fields take as
\begin{equation}
D^{(\alpha)}\equiv R^{(\alpha)\mu} \partial_\mu,\qquad  \alpha=0,1,2,3.
\end{equation}
where
\begin{equation}
D^{(0)}=\frac{\partial}{\partial t},\quad D^{(1)}=-\sin\varphi\frac{\partial}{\partial \theta}-\cot\theta\cos\varphi\frac{\partial}{\partial \varphi},\quad D^{(2)}=\cos\varphi\frac{\partial}{\partial\theta}-\cot\theta\sin\varphi\frac{\partial}{\partial\varphi},\quad D^{(3)}=\frac{\partial}{\partial\varphi}.
\end{equation}
These Killing vector fields describe a time-translation invariance and stationary rotation symmetrical gravitational field.

By using spin-tensor notation Eq.(4) and solving equation Eq.(8), the corresponding Killing scalars is obtained
\begin{eqnarray}
&& B^{(0)}=\frac{\alpha}{2r^2}S^{tr},\nonumber\\
&& B^{(1)}=-r\sin\varphi S^{r\theta}-r\sin\theta\cos\theta\cos\varphi S^{r\varphi}+r^2\sin^2\theta\cos\varphi S^{\theta\varphi},\nonumber\\
&& B^{(2)}=r\cos\varphi S^{r\theta}-r\sin\theta\cos\theta\sin\varphi S^{r\varphi}+r^2\sin^2\theta\sin\varphi S^{\theta\varphi},\nonumber\\
&& B^{(3)}=r\sin^2\theta S^{r\varphi}+r^2\sin\theta\cos\theta S^{\theta\varphi},
\end{eqnarray}
 and the four conserved quantities $J^{(\alpha)}$ can also be found as
\begin{eqnarray}
&& J^{(0)}\equiv E=m\bigg(1-8\pi\eta^2-\frac{\alpha}{r}\bigg)\frac{dt}{d\tau}+\frac{\alpha}{2r^2}S^{tr},\nonumber\\
&& J^{(1)}=-r\sin\varphi\bigg(mr\frac{d\theta}{d\tau}+S^{r\theta}\bigg)-\cos\varphi\bigg(\cot\theta J^{(3)}-r^2S^{\theta\varphi}\bigg),\nonumber\\
&& J^{(2)}=r\cos\varphi\bigg(mr\frac{d\theta}{d\tau}+S^{r\theta}\bigg)-\sin\varphi\bigg(\cot\theta J^{(3)}-r^2S^{\theta\varphi}\bigg),\nonumber\\
&& J^{(3)}=r\sin^2\theta\bigg(mr\frac{d\varphi}{d\tau}+S^{r\varphi}\bigg)+r^2\sin\theta\cos\theta S^{\theta\varphi}.
\end{eqnarray}

Furthermore,  from Eq.(3), covariant constants $\psi^\mu$ are
\begin{eqnarray}
&& \frac{d\psi^t}{d\tau}=-\frac{\alpha}{2r^2}\bigg(1-8\pi\eta^2-\frac{\alpha}{r}\bigg)^{-1}\bigg(\frac{dr}{d\tau}\psi^t+\frac{dt}{d\tau}\psi^r\bigg),\nonumber\\
&& \frac{d\psi^r}{d\tau}=r\bigg(1-8\pi\eta^2-\frac{3\alpha}{2r}\bigg)\bigg(\frac{d\theta}{d\tau}\psi^\theta+\sin^2\theta\frac{d\varphi}{d\tau}\psi^\varphi\bigg),\nonumber\\
&& \frac{d\psi^\theta}{d\tau}=-\frac{1}{r}\bigg(\frac{dr}{d\tau}\psi^\theta+\frac{d\theta}{d\tau}\psi^r\bigg)+\sin\theta\cos\theta\frac{d\varphi}{d\tau}\psi^\varphi,\nonumber\\
&& \frac{d\psi^\varphi}{d\tau}=-\bigg(\frac{1}{r}\frac{dr}{d\tau}+\cot\theta\frac{d\theta}{d\tau}\bigg)\psi^\varphi-\frac{1}{r}\frac{d\varphi}{d\tau}\psi^r-\cot\theta\frac{d\varphi}{d\tau}\psi^\theta.
\end{eqnarray}

Altogether we get twelve equations of motion. Of course, we know that only eight equations, i.e. four velocities and four components of $\psi$, are independent. Considering motion $H=-\frac{m}{2}$, which implies geodesic motion $g_{\mu\nu}{dx^\mu}{dx^\nu}=-d\tau^2$, we obtain
\begin{eqnarray}
\frac{dt}{d\tau}&=&\bigg(1-8\pi\eta^2-\frac{\alpha}{r}\bigg)^{-1}\bigg(\frac{E}{m}-\frac{\alpha}{2mr^2}S^{tr}\bigg),\nonumber\\
\frac{dr}{d\tau}&=&\bigg\{\bigg(1-8\pi\eta^2-\frac{\alpha}{r}\bigg)^2\bigg(\frac{dt}{d\tau}\bigg)^2-\bigg(1-8\pi\eta^2-\frac{\alpha}{r}\bigg) \nonumber\\
& &-r^2\bigg(1-8\pi\eta^2-\frac{\alpha}{r}\bigg)\bigg[\bigg(\frac{d\theta}{d\tau}\bigg)^2+\sin^2\theta\bigg(\frac{d\varphi}{d\tau}\bigg)^2\bigg]\bigg\}^{\frac{1}{2}},\nonumber\\
\frac{d\theta}{d\tau}&=&\frac{1}{mr^2}\bigg(J^{(2)}\cos\varphi-J^{(1)}\sin\varphi-rS^{r\theta}\bigg),\nonumber\\
\frac{d\varphi}{d\tau}&=&\frac{J^{(3)}}{mr^2\sin^2\theta}-\frac{1}{mr}S^{r\varphi}-\frac{1}{m}\cot\theta S^{\theta\varphi}.
\end{eqnarray}
From the independent linear combination of $J^{(1)}$ and $J^{(2)}$, we obtain
\begin{equation}
r^2\sin\theta S^{\theta\varphi}=J^{(1)}\sin\theta\cos\varphi+J^{(2)}\sin\theta\sin\varphi+J^{(3)}\cos\theta,
\end{equation}
this equation implies that there is only the spin angular momentum in the radial direction. From the supersymmetric constraint $Q=0$ (Eq.(14)), we obtain
\begin{equation}
\bigg(1-8\pi\eta^2-\frac{\alpha}{r}\bigg)\frac{dt}{d\tau}\psi^t=\bigg(1-8\pi\eta^2-\frac{\alpha}{r}\bigg)^{-1}\frac{dr}{d\tau}\psi^r+r^2\frac{d\theta}{d\tau}\psi^\theta+r^2\sin^2\theta\frac{d\varphi}{d\tau}\psi^\varphi.
\end{equation}
which implies $\Gamma_\ast=Q^\ast=0$. Expression (23) solves the first part of Eq.(20). The others $S^{ij}$ can be described by
\begin{eqnarray}
&& \frac{dS^{r\theta}}{d\tau}=-\frac{1}{r}\frac{dr}{d\tau}S^{r\theta}+\sin\theta\cos\theta\frac{d\varphi}{d\tau}S^{r\varphi}-r\sin^2\theta\bigg(1-8\pi\eta^2-\frac{3\alpha}{2r}\bigg)\frac{d\varphi}{d\tau}S^{\theta\varphi},\nonumber\\
&& \frac{dS^{r\varphi}}{d\tau}=\cot\theta\frac{d\varphi}{d\tau}S^{r\theta}-\bigg(\frac{1}{r}\frac{dr}{d\tau}+\cot\theta\frac{d\theta}{d\tau}\bigg)S^{r\varphi}+r\bigg(1-8\pi\eta^2-\frac{3\alpha}{2r}\bigg)\frac{d\theta}{d\tau}S^{\theta\varphi},\nonumber\\
&& \frac{dS^{\theta\varphi}}{d\tau}=\frac{1}{r}\frac{d\varphi}{d\tau}S^{r\theta}-\frac{1}{r}\frac{d\theta}{d\tau}S^{r\varphi}-\bigg(\frac{2}{r}\frac{dr}{d\tau}+\cot\theta\frac{d\theta}{d\tau}\bigg)S^{\theta\varphi}.
\end{eqnarray}

The equation of $S^{\theta\varphi}$ is solved by Eq.(22). Eq.(23) allows us to rewrite the time-like components $S^{it}$ in terms of the space-like $S^{ij}$. Using Eq.(23), the first part of Eq.(21) can be written as follows
\begin{equation}
\frac{dt}{d\tau}=\bigg(1-8\pi\eta^2-\frac{\alpha}{r}\bigg)^{-1}\bigg[\frac{E}{m}+\frac{\alpha}{2E}\bigg(\frac{d\theta}{d\tau}S^{r\theta}+\sin^2\theta\frac{d\varphi}{d\tau}S^{r\varphi}\bigg)\bigg].
\end{equation}

By integrating Eqs.(21)(excluding the first line), (24) and (25), we can solve the equation of motion for the coordinates and spins. For $\eta=0$, these equations correspond to those in Schwarzschild spacetime\cite{10}.

\section{Critical escape energy and velocity of the spinning particle}
 For simplifying the motion, we choose $\theta=\frac{\pi}{2}$. The angular momentum of the scalar particle is conserved. But it is not the generic case of the spinning particle, for which only the total angular momentum is conserved. It occurs only in two situations: (i) the orbital angular momentum vanishes, or  (ii) spin and orbital angular momentum are parallel.

With the condition $\theta=\frac{\pi}{2}$ and $\dot{\theta}=0$, the equations of motion Eq.(21) and Eq.(24) become as follows
\begin{eqnarray}
&& \frac{dt}{d\tau}=\bigg(1-8\pi\eta^2-\frac{\alpha}{r}\bigg)^{-1}\bigg(\frac{E}{m}+\frac{\alpha}{2E}\frac{d\varphi}{d\tau}S^{r\varphi}\bigg),\nonumber\\
&& \frac{dr}{d\tau}=\bigg[\bigg(1-8\pi\eta^2-\frac{\alpha}{r}\bigg)^2\bigg(\frac{dt}{d\tau}\bigg)^2-\bigg(1-8\pi\eta^2-\frac{\alpha}{r}\bigg)-r^2\bigg(1-8\pi\eta^2-\frac{\alpha}{r}\bigg)\bigg(\frac{d\varphi}{d\tau}\bigg)^2\bigg]^{\frac{1}{2}},\nonumber\\
&& \frac{d\varphi}{d\tau}=\frac{1}{mr^2}J^{(3)}-\frac{1}{mr}S^{r\varphi},\nonumber\\
&& \frac{d}{d\tau}(rS^{r\theta})=-\bigg(1-8\pi\eta^2-\frac{3\alpha}{2r}\bigg)r^2S^{\theta\varphi}\frac{d\varphi}{d\tau},\nonumber\\
&& \frac{d}{d\tau}(rS^{r\varphi})=0.
\end{eqnarray}

The third and the last parts of Eq.(26) means that the orbital angular momentum and the component of the spin, perpendicular to the particle moving plane, are conserved
\begin{equation}
rS^{r\varphi}=\Sigma, \qquad mr^2\frac{d\varphi}{d\tau}=J^{(3)}-\Sigma=L.
\end{equation}
where $\Sigma$ and $L$ are two constants. The gravitational red-shift which is defined by the first part of Eq.(26) takes as follows
\begin{equation}
dt=\frac{d\tau}{1-8\pi\eta^2-\frac{\alpha}{r}}\bigg(\frac{E}{m}+\frac{\alpha}{2mEr^3}L\Sigma\bigg).
\end{equation}
For $L\neq0$, the time-dilation receives a contribution from spin-orbit coupling. This implies that time-dilation is not only a purely geometric effect, but also a dynamical component\cite{13}. The third part of Eq.(21), Eq.(22), and the fourth part of Eq.(26) with $\theta=\frac{\pi}{2}$, give two possibilities for planar motion:
\begin{equation}
(1)\dot{\varphi}=0 \qquad(2)S^{\theta\varphi}=0.
\end{equation}

In the situation (1), $\dot{\varphi}=0$ implies the orbital angular momentum vanishes, $L=mr^2\dot{\varphi}=0$. The particle moves along a fixed radius. The equation of motion about the spinless particle for a distant observer is described by
\begin{equation}
\frac{dr}{dt}=\bigg(1-8\pi\eta^2-\frac{\alpha}{r}\bigg)\sqrt{1-\frac{m^2}{E^2}\bigg(1-8\pi\eta^2-\frac{\alpha}{r}\bigg)},
\end{equation}
as in the case of a spinless particle, if we choose $\varphi=0$, we find the spin tensor components are all conserved
\begin{equation}
r^2S^{\theta\varphi}=J^{(1)}, \qquad rS^{r\theta}=J^{(2)}, \qquad rS^{r\varphi}=J^{(3)}.
\end{equation}

In the situation (2), $\dot{\varphi}\neq0$ implies that
\begin{equation}
S^{\theta\varphi}=0,\qquad S^{r\theta}=0,\qquad J^{(1)}=J^{(2)}=0.
\end{equation}
which means the spin is parallel to the orbital angular momentum. For $\dot{r}$ and $\dot{\varphi}$, Eq.(26) gives the particle orbital equation
\begin{equation}
\frac{1}{r^2}\bigg(\frac{dr}{d\varphi}\bigg)^2=\frac{[E^2-m^2(1-8\pi\eta^2)]r^2}{L^2}-(1-8\pi\eta^2)+\frac{m\alpha}{L}\bigg(\frac{mr}{L}+\frac{J^{(3)}}{mr}\bigg).
\end{equation}

Using the dimensionless variables
\begin{equation}
\epsilon=\frac{E}{m},\qquad x=\frac{r}{\alpha},\qquad J=\frac{J^{(3)}}{m\alpha},\qquad \Xi=\frac{rS^{r\varphi}}{m\alpha},\qquad l=\frac{L}{m\alpha}=J-\Xi,\qquad \Delta=\frac{\Sigma}{L}=\frac{\Xi}{J-\Xi},
\end{equation}
we obtain
\begin{equation}
\frac{l^2}{x^4}\bigg(\frac{dx}{d\varphi}\bigg)^2=\alpha^2\dot{x}^2=\epsilon^2-U_R(x,l^2),
\end{equation}
where
\begin{equation}
U_{eff}(x,l^2)=(1-8\pi\eta^2)-\frac{1}{x}+(1-8\pi\eta^2)\frac{l^2}{x^2}-\frac{l^2(1+\Delta)}{x^3}
\end{equation}
defines an effective potential. $\Delta\ll1$ describes a realistic physics situation.

For bound state orbits, it is necessary that $\epsilon<1$. The function $U_{eff}(x,l^2)$ has a point of inflection which corresponds to a circular orbit with minimum radius
\begin{equation}
x=(1-8\pi\eta^2)l^2=\frac{3(1+\Delta)}{1-8\pi\eta^2}.
\end{equation}
The energy for this critical orbit is given by
\begin{equation}
\epsilon^2_{crit}=\frac{(1-8\pi\eta^2)(8+\Delta)}{9},
\end{equation}
and the time-dilation factor is expressed by
\begin{equation}
\bigg(\frac{dt}{d\tau}\bigg)_{crit}=\frac{1}{1-8\pi\eta^2-\frac{1}{x}}\bigg(\epsilon_{crit}+\frac{l^2\Delta}{2\epsilon_{crit}x^3}\bigg).
\end{equation}
With $\eta=0$, Eqs.(37-39) reduce to the Schwarzschild results.

The orbit of the particle which approaches precessing ellipse (because of relativistic effects) is given by
\begin{equation}
x=\frac{\kappa}{1+e\cos[\varphi-\omega(\varphi)]},
\end{equation}
here $\kappa=\frac{k}{\alpha}$ with $k$ the semilatus rectum and $e$ is the eccentricity with $0<e<1$. The perihelion and aphelion are defined by
\begin{equation}
\varphi^{(k)}_{ph}-\omega(\varphi^{(k)}_{ph})=2k\pi, \qquad \varphi^{(k)}_{ah}-\omega(\varphi^{(k)}_{ah})=(2k+1)\pi.
\end{equation}
The angle $\varphi^{(k)}_{ph}$ is the $k$-th perihelion of the particle, while $\omega(\varphi^{(k)}_{ph})$ is the amount of precession of the perihelion after $k$ revolutions. Hence the precession of the perihelion after one revolution is
\begin{equation}
\Delta\omega=\omega(\varphi^{(1)}_{ph})-\omega(\varphi^{(0)}_{ph})=\varphi^{(1)}_{ph}-\varphi^{(0)}_{ph}-2\pi\equiv\Delta\varphi-2\pi.
\end{equation}
The $\epsilon$ is a constant of motion and at the perihelion/aphelion is given as follows
\begin{equation}
\epsilon^2=(1-8\pi\eta^2)-\frac{1\pm e}{\kappa}+(1-8\pi\eta^2)l^2\bigg(\frac{1\pm e}{\kappa}\bigg)^2-l^2(1+\Delta)\bigg(\frac{1\pm e}{\kappa}\bigg)^3.
\end{equation}
Comparing both expressions for $\epsilon^2$, we obtain
\begin{equation}
l^2=\frac{\kappa^2}{2\kappa(1-8\pi\eta^2)-(1+\Delta)(3+e^2)}.
\end{equation}
Using above results and defining
\begin{equation}
y=\varphi-\omega(\varphi),
\end{equation}
we rewrite Eq. (33) in the form
\begin{equation}
d\varphi=\frac{dy}{\sqrt{(1-8\pi\eta^2)-\frac{1+\Delta}{\kappa}(3+e\cos y)}}.
\end{equation}
We find $\Delta\varphi$ in Eq.(42), can be obtained by integrating Eq.(46) from one perihelion to the next one with $0\leq y\leq 2\pi$
\begin{equation}
\Delta\varphi=\frac{1}{\sqrt{a}}\int^{2\pi}_0\frac{dy}{\sqrt{1-\frac{b}{a}\cos y}},
\end{equation}
where
\begin{equation}
a=(1-8\pi\eta^2)-3F, \qquad b=eF, \qquad F=\frac{1+\Delta}{\kappa}.
\end{equation}
Integrating Eq.(47), we obtain the following expression for $\Delta\varphi$
\begin{eqnarray}
\Delta\varphi =\frac{2\pi}{(1-8\pi\eta^2)^{\frac{5}{2}}}\bigg[(1-8\pi\eta^2)^2+\frac{3M}{k}(1-8\pi\eta^2)(1+\Delta)+\frac{3M^2}{4k^2}(e^2+18)(1+\Delta)^2+\cdots\bigg].
\end{eqnarray}
For $\Delta=0$, the second term in this expression is the lowest-order contribution to the relativistic precession of the perihelion. We find the spin of a particle contributes to the lowest-order precession can be keeping terms of first order in $\Delta$. For $\eta=0$, the result in Eq.(49) corresponds to that in Schwarzschild spacetime\cite{10}.

Then we will discuss the motion of a spinning particle originally revolving around the black hole in the innermost stable circular orbit kick by another particle or photon according to the procedure given in Ref.\cite{14}. When a spinning particle moves on a circular orbit in the equatorial plane, we take the particle spin vector perpendicular to this plane. Thus we set\cite{15,16}
\begin{equation}
S^\mu=\frac{s}{r}(0,0,1,0),
\end{equation}
and the spin tensor is related to its spin vector by\cite{16,17}
\begin{equation}
S^{\mu\nu}=\frac{\epsilon^{\mu\nu\alpha\beta}}{m\sqrt{-g}}p_\alpha S_\beta
\end{equation}
where  $\epsilon^{\alpha\beta\mu\nu}$ is the alternating symbol, which implies that the spin tensor has only one independent component. After the collision, the particle will move in a new plane tilted to the original equatorial plane. The motion of the particle has two types: bounded motion or escape to infinity. The result is determined by the detail of the collision mechanism. For large values of $\epsilon-\epsilon_0$, the particle will leave away from the original equatorial plane and escape to infinity. Here $\epsilon$ is the energy after collision and $\epsilon_0$ is the energy before collision which is given by Eq.(38). In order to simply the problem, we impose the restrictions: (i) the $z$ component of the angular momentum$(J^{(3)})$ is not changed, (ii) the initial radial velocity after the collision remains the same, and (iii) the spin vector and the component of the spin perpendicular to the particle motion plane are not changed. Under these restrictions, only the parameter $\epsilon$ determine the motion of the spinning particle. So after collision, the particle requires a velocity $v_\perp=-r\dot{\theta}_0$ which is in the direction orthogonal to the equatorial plane. That's to say, the kick only gives the particle a velocity $v_\perp$.

After the collision, the energy of the particle is
\begin{equation}
\epsilon=\sqrt{\epsilon^2_0+(1-8\pi\eta^2-\frac{\alpha}{r})v^2_\bot}.
\end{equation}

The spinning particle moving in the gravitational field obeys the equation
\begin{equation}
\frac{D^2x^\mu}{D\tau^2}=\frac{1}{2}S^{\kappa\lambda} R^\mu_{\ \nu\kappa\lambda}\dot{x}^\nu,
\end{equation}
which is the first part of Eq.(3). $\dot{x}^\mu=u^\mu$ is the particle's four-velocity, $u^\mu u_\mu=-1$. Using this normalization condition and the first and last part of Eq.(21), the $x$ and $\theta$ components of Eq.(53) and $\epsilon$ can be written in dimensionless variables (given by Eq.(34) and $\sigma=\frac{\tau}{\alpha}$) as follows
\begin{eqnarray}
&& \ddot{x}=-\frac{1}{2x^2}+x\bigg(1-8\pi\eta^2-\frac{3}{2x}\bigg)\bigg(\frac{d\theta}{d\sigma}\bigg)^2+\bigg(1-8\pi\eta^2-\frac{3}{2x}\bigg)\bigg(\frac{J^2}{x^3\sin^2\theta}-\frac{2J\Xi}{x^3}\bigg)+\frac{3J\Xi}{4x^4},\\
&& \ddot{\theta}=-\frac{2}{x}\bigg(\frac{dx}{d\sigma}\bigg)\bigg(\frac{d\theta}{d\sigma}\bigg)+\frac{J^2\cos\theta}{x^4\sin^3\theta}-\frac{2J\Xi\cos\theta}{x^4\sin\theta},\\
&& \epsilon^2=\bigg(\frac{dx}{d\sigma}\bigg)^2+x^2\bigg(1-8\pi\eta^2-\frac{1}{x}\bigg)\bigg(\frac{d\theta}{d\sigma}\bigg)^2+U_{eff},\\
&& U_{eff}=\bigg(1-8\pi\eta^2-\frac{1}{x}\bigg)\bigg(1+\frac{J^2}{x^2\sin^2\theta}-\frac{2J\Xi}{x^2}\bigg)-\frac{J\Xi}{x^3}.
\end{eqnarray}

The energy of a spinning particle revolving around the black hole in the innermost stable circular orbit of the equatorial plane is
\begin{equation}
\epsilon_0=\sqrt{\bigg(1-8\pi\eta^2-\frac{1}{x_0}\bigg)\bigg(1+\frac{J^2}{x^2_0}-\frac{2J\Xi}{x^2_0}\bigg)-\frac{J\Xi}{x^3_0}},
\end{equation}
which corresponds to Eq.(38). When $v_\bot>0$, the energy of the particle changes from $\epsilon_0$ to
\begin{equation}
\epsilon=\sqrt{\epsilon^2_0+(1-8\pi\eta^2-\frac{1}{x_0})v^2_\bot}.
\end{equation}
Using Eq.(37), the parameters $J$ and $\Xi$ can be defined as
\begin{eqnarray}
&& J=\pm\sqrt{\frac{(1-8\pi\eta^2)x^3_0}{9}},\\
&& \Xi=\pm\bigg[1-\frac{3}{(1-8\pi\eta^2)x_0}\bigg]\sqrt{\frac{(1-8\pi\eta^2)x^3_0}{9}},
\end{eqnarray}
where $x_0=\frac{3(1+\Delta)}{1-8\pi\eta^2}$, which is determined by $\Delta$ and $8\pi\eta^2$, is the radius of the orbit. The sign $\pm$ show the directions of the $z$ component of the angular momentum which do not influence the particle motion. $J$ and $\Xi$ are specified by $8\pi\eta^2$ and $\Delta$. With fixed $8\pi\eta^2$ and $\Delta$, only $\epsilon$ serves to the motion of the kicked particle.

Given the value of $8\pi\eta^2$, $\Delta$ and the initial energy $\epsilon$ of the kicked spinning particle, we can integrate Eqs.(54) and (55) numerically. We simulated the trajectory corresponding to the initial condition. The Eqs.(56) and (58) are used to check independently for numerical precision. The numerical integration shows that there are three types of final motion for the particle:

(1) The spinning particle is on a bounded  trajectory around the black hole.

(2) The spinning particle escapes to infinity $x\cos\theta\rightarrow\infty$.

(3) The spinning particle escapes to infinity $x\cos\theta\rightarrow-\infty$.

The particle is considered to escape to infinite if $\mid x\cos\theta\mid$ reaches $10^3$, and the integrate computation time was chosen as $\sigma=10^6$.

 \begin{figure}[!htb]
                \centering
                \subfigure[]{         
                \label{}         
                \includegraphics[width=0.45\textwidth ]{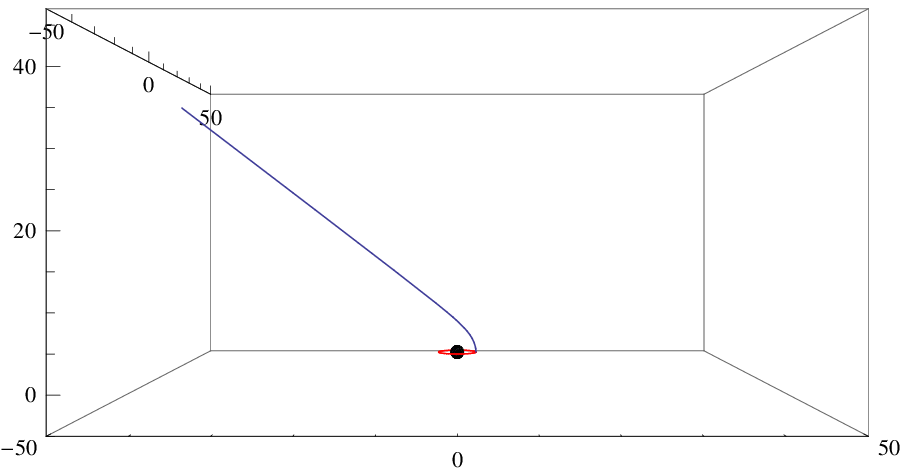}}
                \subfigure[]{         
                \label{}         
                \includegraphics[width=0.45\textwidth ]{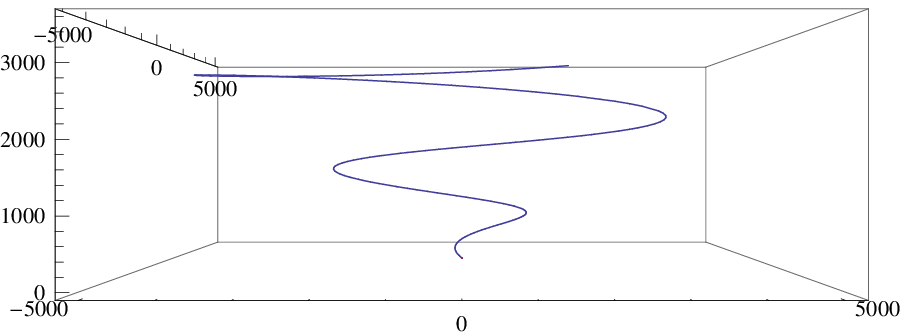}}
                \subfigure[]{         
                \label{}         
                \includegraphics[width=0.45\textwidth ]{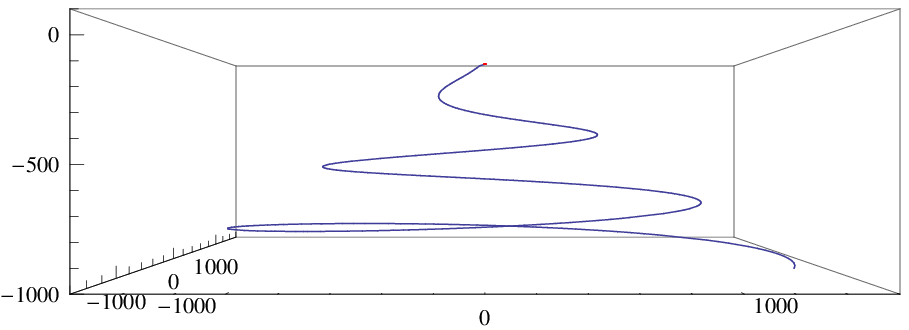}}
                 \subfigure[]{         
                \label{}         
                \includegraphics[width=0.45\textwidth ]{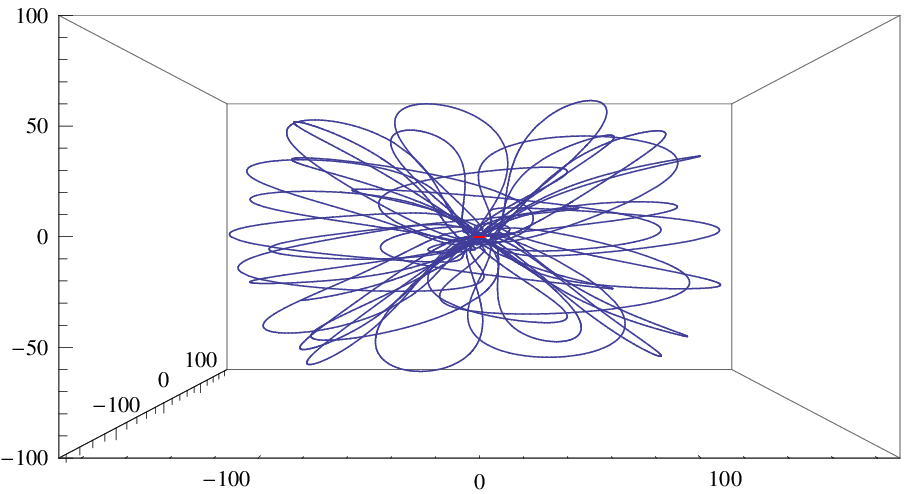}}
                \caption{(Color online) These figures show the particle trajectories for different initial conditions. With fixed $8\pi\eta^2=0.01$: (a) $\Delta=0,\epsilon=1.247$, (b) $\Delta=0.01,\epsilon=1.247$, (c) $\Delta=0.01,\epsilon=0.994$, (d) $\Delta=0.01,\epsilon=0.990$.}         
                \label{}         
        \end{figure}

         \begin{figure}[!htb]
                \centering
                \subfigure[]{         
                \label{}         
                \includegraphics[width=0.47\textwidth ]{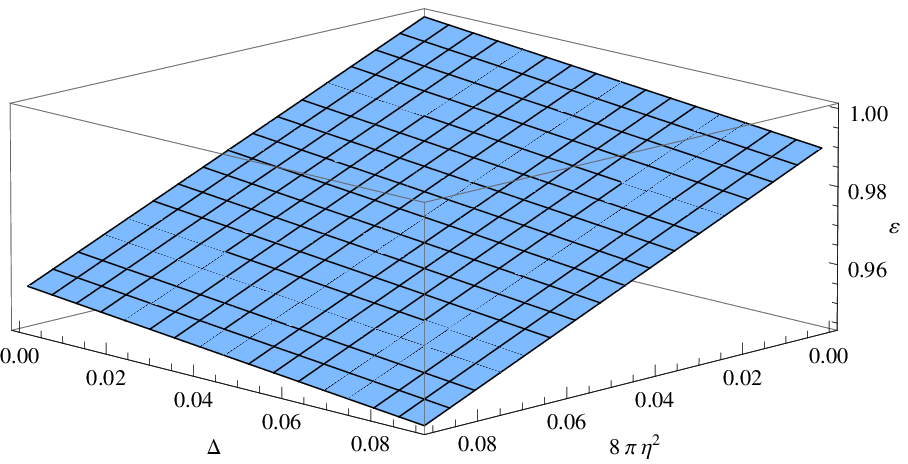}}
                \subfigure[]{         
                \label{}         
                \includegraphics[width=0.47\textwidth ]{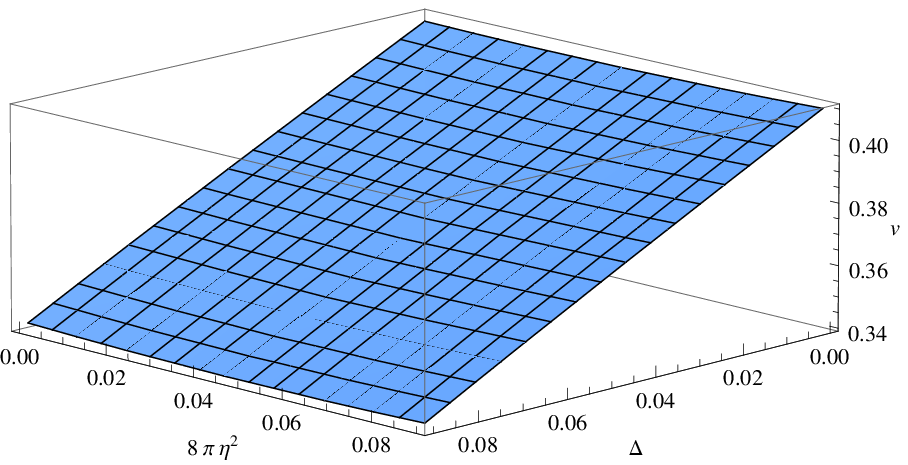}}
                \caption{(Color online) These figures show the critical escape energy and velocity of the spinning particle, respectively.}         
                \label{}         
        \end{figure}

Fig.1 shows the trajectories of the particle for different initial conditions. Fig.1(a) shows the escape trajectory for the spinless particle, and the others give the trajectory for the spinning particle: Fig.1(b) shows the particle escapes to $x\cos\theta\rightarrow\infty$, Fig.1(c) shows the particle escapes to $x\cos\theta\rightarrow-\infty$, Fig.1(d) shows the bounded trajectory of the particle. From these figures, we can find that the escape trajectory for the spinning particle is the cylindro-conical helix which is different form the spinless particle. The trajectory shows that the black hole is surrounded by the compressed spherical particle cloud which is formed by the bounded spinning particle.

From the result of the numerical integration, we obtain the relation of the critical escape energy $\varepsilon$, $8\pi\eta^2$ and $\Delta$ which is shown in Table 1 and Fig.2(a). Different from the result in Ref.\cite{14}, we have not found the chaos in the critical escape domain. At the same time Fig.2(a) and Table 1 illustrate that a larger $\eta$ or $\Delta$ will require a smaller critical escape energy $\varepsilon$.

\begin{table}

\caption{The relation among critical escape energy $\varepsilon$, $\Delta$ and $8\pi\eta^2$.}
\begin{center}
 \begin{tabular}{l|c}
  \hline
 {\backslashbox{$8\pi\eta^2$}{$\Delta$}}&{0.000 0.010 0.020 0.030 0.040 0.050 0.060 0.070 0.080 0.090}\\
  \hline
 0.000&{1.000 0.999 0.998 0.997 0.995 0.994 0.993 0.992 0.991 0.990}\\
 0.010&{0.995 0.994 0.993 0.992	0.990 0.989	0.988 0.987	0.986 0.985}\\
 0.020&{0.990 0.989	0.988 0.987	0.985 0.984	0.983 0.982	0.981 0.980}\\
 0.030&{0.985 0.984	0.983 0.982	0.980 0.979	0.978 0.977	0.976 0.975}\\
 0.040&{0.980 0.979	0.978 0.977	0.975 0.974	0.973 0.972	0.971 0.970}\\
 0.050&{0.975 0.974	0.973 0.971	0.970 0.969	0.968 0.967	0.966 0.965}\\
 0.060&{0.970 0.969	0.967 0.966	0.965 0.964	0.963 0.962	0.961 0.960}\\
 0.070&{0.965 0.964	0.962 0.961	0.960 0.959	0.958 0.957	0.956 0.955}\\
 0.080&{0.960 0.958	0.957 0.956	0.955 0.954	0.953 0.952	0.950 0.949}\\
 0.090&{0.954 0.953	0.952 0.951	0.950 0.949	0.947 0.946	0.945 0.944}\\
  \end{tabular}
\end{center}
   \end{table}

Using the numerical results, we can estimate the critical escape energy of the particle. The approximate analytical expression is
\begin{equation}
\varepsilon\approx1+\frac{175.2604(-0.0017+\Delta)+778.1768(8\pi\eta^2)}{(\Delta-68.9565)(22.2024+\Delta)-(8\pi\eta^2)}.
\end{equation}
The critical escape velocity can be obtained by Eq.(37), Eq.(59) and Eq.(62). Fig.2(a) and 2(b) illustrate the critical escape energy and velocity, respectively.

\section{Summary}

In this paper by studied the spinning particle motion in the gravitational field of black hole involving a global monopole. we have derived the constants and the equations of motion and investigate bound state orbits in a plane. We also have derived the conditions that a spinning particle, originally revolving around the black hole in the innermost stable circular orbit, can escape to infinity after being kicked by another particle or photon. We have found that the escape trajectory for the kicked spinning particle is the cylindro-conical helix which is different form the spinless particle, and the trajectory of the spinning particle in the jet exists three different types: bounded trajectory, and escape to infinity $\mid x\cos\theta\mid\rightarrow\infty$. We have obtained the critical escape energy and velocity of the spinning particle. A larger $\eta$ or $\Delta$ will require a smaller critical escape energy $\varepsilon$. It is interesting to analyze the critical escape phenomenon of the charged spinning particle in the gravitational field with magnetic field.

\section{Acknowledgments}

The work was supported by the State Key Development Program for Basic Research of China (Grant No. 2010CB832800), the National Natural Science Foundation of China (Grant No. 10873004).


\end{CJK*}  

\begin{thebibliography}{99}
\bibitem{Berezin} Berezin F A and Marinov M S, {\it Ann. Phys. N.Y.} {\bf 104} 336 (1977)
\bibitem{2} Casalbuoni R, {\it Phys. Lett.} {\bf 62B} 49 (1976)
\bibitem{3} Barducci A, Casalbuoni R and Lusanna L, {\it Nuovo Cimento A} {\bf 35} 377 (1976)
\bibitem{4} Brink L, Deser S, Zumino B, Di Vecchia P and Howe P, {\it Phys. Lett.} {\bf B64} 435 (1976)
\bibitem{5} Brink L, Di Vecchia P and Howe P, {\it Nucl. Phys. B} {\bf 118} 76 (1977)
\bibitem{6} Rietdijk R H and van Holten J W, {\it Class. Quantum Grav.} {\bf 7} 247 (1990)
\bibitem{7} van Holten J W and Rietdijk R H, {\it Journal of Geometry and Physics} {\bf 11} 559 (1993)
\bibitem{8} Rietdijk R H and van Holten J W, {\it Class. Quantum Grav.} {\bf 10} 575 (1993)
\bibitem{9} Gibbons G W, Rietdijk R H and van Holten J W, {\it Nucl. Phys. B} {\bf 404} 42 (1993)
\bibitem{f1} Papapetrou A, {\it Proc. R. Soc. London A} {\bf 209} 248 (1951)
\bibitem{f2} Dixon W G, {\it Proc. R. Soc. London A} {\bf 314} 499 (1970)
\bibitem{f3} van Holten J W, {\it Proc. Seminar 1986~1987 Mathematical Structures in Field Theories}(CWI, Amsterdam,1990) p.109
\bibitem{f4} Souriau J M, {\it Ann. Inst. Henri Poincar\'{e} A} {\bf 20} 315 (1974)
\bibitem{f5} Mohseni M, {\it Int. J. Mod. Phys. D} {\bf 15} 121 (2006)
\bibitem{10} Rietdijk R H and van Holten J W, {\it Class. Quantum Grav.} {\bf 10} 575 (1993)
\bibitem{g1} Suzuki S and Maeda K, {\it Phys. Rev. D} {\bf 55} 4848 (1997)
\bibitem{g2} Verhaaren C and Hirschmann E W, {\it Phys. Rev. D} {\bf 81} 124034 (2010)
\bibitem{g3} Han W, {\it Gen. Relativ. Gravit} {\bf 40} 1831 (2008)
\bibitem{g4} Hossain A M, {\it Gen. Relativ. Gravit} {\bf 35} 285 (2003)
\bibitem{chen1} Chen J H and Wang Y J, {\it Int. J. Mod. Phys. A} {\bf 25} 1439 (2010)
\bibitem{chen2} Zhou S, Chen J H and Wang Y J, {\it Int. J. Mod. Phys. D} {\bf 21} 1250077 (2012)
\bibitem{ww} Wang Y and Wu X, {\it Class. Quantum Grav.} {\bf 28} 025010 (2011)
\bibitem{wx1} Wu X and Xie Y, {\it Phys. Rev. D} {\bf 77} 103012 (2008)
\bibitem{wx2} Wu X and Xie Y, {\it Phys. Rev. D} {\bf 81} 084045 (2010)
\bibitem{zw} Zhong S Y and Wu X, {\it Phys. Rev. D} {\bf 81} 104037 (2010)
\bibitem{14} Al Zahrani A M, Frolov V P and Shoom A A, {\it Phys. Rev. D} {\bf 87} 084043 (2013)
\bibitem{a1} Tamaki T and Sakai N, {\it Phys. Rev. D} {\bf 69} 044018 (2004)
\bibitem{a2} Yamaguchi M, {\it Phys. Rev. D} {\bf 64} 081301 (2001)
\bibitem{a3} Jiang Q Q and Wu S Q, {\it Phys. Lett. B} {\bf 635} 151 (2006)
\bibitem{a4} Peng J J and Wu S Q, {\it Chin. Phys.} {\bf 17} 825 (2007)
\bibitem{a5} Watabe H and Torii T, {\it JCAP} {\bf 02} 001 (2004)
\bibitem{11} Rietdijk R H and van Holten J W, {\it Class. Quantum Grav.} {\bf 7} 247 (1990)
\bibitem{12} Barriola M and Vilenkin A, {\it Phys. Rev. Lett.} {\bf 63} 341 (1989)
\bibitem{13} van Holten J W, {\it Phys. A} {\bf 182} 279 (1992)
\bibitem{15} Mortazavimanesh M and Mohseni M, {\it Gen. Relativ. Gravit} {\bf 41} 2697 (2009)
\bibitem{16} Mohseni M, {\it Gen. Relativ. Gravit} {\bf 42} 2477 (2010)
\bibitem{17} Tod K P, de Felice F and Calvani M, {\it Nuovo Cimento B} {\bf 34} 365 (1976)


\end{thebibliography}
\end{document}